\documentclass[reprint,preprintnumbers,
superscriptaddress,
amsmath,amssymb,aps]{revtex4-2}
\usepackage[utf8]{inputenc}

\usepackage{graphicx}
\usepackage{dcolumn}
\usepackage{bm}

\usepackage{color}
\usepackage[dvipsnames, table]{xcolor}
\definecolor{blue}{rgb}{0,0,0.5}
\usepackage[colorlinks,linkcolor=blue,urlcolor=blue,citecolor=blue]{hyperref}
\usepackage{graphicx}
\usepackage{slashed}
\usepackage{enumitem}
\usepackage{bm}
\usepackage{physics}
\usepackage{xspace}
\usepackage[T1]{fontenc}
\usepackage{lmodern}
   
\usepackage{mathrsfs}

\usepackage{float} 

\newcommand{\be}{\begin{equation}}
\newcommand{\ee}{\end{equation}}
\newcommand{\bea}{\begin{eqnarray}}
\newcommand{\eea}{\end{eqnarray}}

\newcommand{\mc}{\mathcal}

\newcommand{\noi}{\noindent}

\newcommand{\miss}{{\rm miss}}

\newcommand{\Mrecoil}{M_{\rm recoil}}
\newcommand{\Mreco}{M_{\rm recoil}}

\def\T{{\rm T}}

\def\MT2{M_{\rm T2}}
\def\M2{M_{\rm 2}}
\def\k#1{{\boldsymbol{k}}_{#1}}

\def\p#1{{\boldsymbol{p}}_{#1}}

\def\true{{\rm true}}
\def\max{{\rm max}}
\def\min{{\rm min}}

\def\sig{{\rm sig}}

\newcommand{\GeV}{{\rm GeV}}

\newcommand{\mvv}{m_{\nu \bar \nu}}
\newcommand{\mvva}{\hat m_{\nu \bar \nu}}

\DeclareOldFontCommand{\rm}{\normalfont\rmfamily}{\mathrm}
\DeclareOldFontCommand{\sf}{\normalfont\sffamily}{\mathsf}
\DeclareOldFontCommand{\tt}{\normalfont\ttfamily}{\mathtt}
\DeclareOldFontCommand{\bf}{\normalfont\bfseries}{\mathbf}
\DeclareOldFontCommand{\it}{\normalfont\itshape}{\mathit}
\DeclareOldFontCommand{\sl}{\normalfont\slshape}{\@nomath\sl}
\DeclareOldFontCommand{\sc}{\normalfont\scshape}{\@nomath\sc}

\begin{document}

\preprint{LAPTH-049/22, CTPU-PTC-22-20}

\title{\boldmath A new approach to semi-leptonic tags in $B$-meson semi-invisible decays}

\author{Gaetano de Marino}

\email{demarino@ijclab.in2p3.fr}

\affiliation{%
{\itshape Universit\'e Paris-Saclay, CNRS/IN2P3, IJCLab, 91405 Orsay, France}
}%

\author{Diego Guadagnoli}

\email{diego.guadagnoli@lapth.cnrs.fr}

\affiliation{%
{\itshape LAPTh, Universit\'{e} Savoie Mont-Blanc et CNRS, 74941 Annecy, France}
}%

\author{Chan Beom Park}

\email{cbpark@jnu.ac.kr}

\affiliation{%
{\itshape Center for Theoretical Physics of the Universe, Institute for Basic Science (IBS), 34126 Daejeon, Korea}
}%

\affiliation{%
{\itshape Department of Physics, Chonnam National University, Gwangju 61186, Korea}
}%

\author{Karim Trabelsi}

\email{karim.trabelsi@in2p3.fr}

\affiliation{%
{\itshape Universit\'e Paris-Saclay, CNRS/IN2P3, IJCLab, 91405 Orsay, France}
}%

\begin{abstract}
\noi Kinematic variables designed for pairwise decays to partly undetected final states---a prominent example being $\MT2$ and its Lorentz-invariant version $\M2$---have been extensively deployed in high-$\p{\T}$ collider searches. A new range of potential applications at flavour facilities---where $B$ mesons or $\tau$ leptons are also pairwise produced---was recently proposed.

One general challenge in these decays arises if both the signal parent and the `other' parent, often used as a tag, decay semi-invisibly. In such cases, which notably include semi-leptonic tags, signal identification is generally hindered by the ensuing lack of knowledge of the signal-parent boost. $\M2$ helps precisely to overcome this challenge, and allows to leverage the otherwise superior efficiency of semi-leptonic decays.

Our strategy rests on two novel constraints that can be imposed on $M_2$. The first is that of the known mass of the decaying-parent mass squared which, in connection with other constraints, gives rise to $M_{2sB}$.
The second is on the flight direction of the signal parent, often well reconstructed at facilities with high vertexing capabilities such as Belle II and LHCb. This constraint gives rise to the $M_{2V}$ variable, that can be used even at facilities where the collision energy is not known.

We test these ideas in a decay of great current interest in the context of the persistent discrepancies in $B$ decays, namely $B \to K \tau \mu$. We find that a bare-bones application of $M_{2sB}$ leads, alone, to an improvement that is already halfway between the current approach and the ``truth-level'' semi-leptonic case. Ceteris paribus---in particular statistics---our approach thus makes semi-leptonic tags competitive with fully reconstructed hadronic tags.
\end{abstract}

\maketitle

\newcommand{\Btag}{B_{\rm tag}}
\newcommand{\Bsig}{B_{\rm sig}}
\newcommand{\Ktag}{K_{\rm tag}}
\newcommand{\Ksig}{K_{\rm sig}}
\newcommand{\lsig}{\ell_{\rm sig}}
\newcommand{\ltag}{\ell_{\rm tag}}
\newcommand{\ltau}{\ell_{\tau}}
\newcommand{\Tag}{{\rm tag}}
\newcommand{\Sig}{{\rm sig}}

\noi Pair-produced particles decaying to partly invisible final states are an ubiquitous topology at high-intensity experiments such as LHCb and Belle II. A prototype example are $B - \bar B$ pairs, each $B$ decaying to final states including at least one neutrino. Of these $B$ mesons, one (`$\Bsig$') produces the `signal' decay, and the other (`$\Btag$') may be used as a `tag'.
Common tag decays are modes that can be fully reconstructed, as is the case for many hadronic tags. This way, even if the signal includes elusive particles, the number of constraints is large enough to close the kinematics. Such strategy, somewhat by definition, excludes from consideration tag decays that contain undetected particles, for instance most semi-leptonic tags. This is unfortunate, because semi-leptonic tags are often clean, thus affording high efficiencies, and most importantly they have large branching fractions---almost 20\% 
for only four $D^{(*)}\ell \nu$ modes.

Precisely events where both $\Bsig$ and $\Btag$ decay semi-invisibly lend themselves to the use of variables such as $\M2$, as recently pointed out in Ref. \cite{Guadagnoli:2021fcj}. Even in the absence of enough constraints to close the event kinematics, these variables include a `built-in' way of estimating the {\em separate} invisible momenta for the two decay chains. This is provided by the so-called $\M2$-Assisted On-Shell invisible momenta, usually referred to as MAOS momenta \cite{Cho:2008tj,Park:2011uz} (see also \cite{Cho:2014naa,  Kim:2017awi} for the specific $\M2$ case). In this context, it was observed that the larger the number of mass constraints used, the more MAOS momenta approach the true momenta.

In this letter we introduce a strategy to exploit the above ideas towards a full event reconstruction in semi-leptonic tags, with the ultimate objective of making them competitive, in resolution, with hadronic tags, and thereby substantially increase the usable statistics in, potentially, any semi-invisible $B$-meson decay.

Our strategy is based on two new kinematic requirements, or `constraints' in the following, that can be enforced on the $\M2$ definition. These constraints are new in the following respects: {\em (i)} they `overload' $\M2$, i.e. they reduce to zero the number of kinematic degrees of freedom in the $\M2$ distribution. Then $\M2$ is no more a distribution---its minimization becomes equivalent to finding the unique solution of the event's kinematic equations; {\em (ii)} we consider kinematic requirements other than on-shell mass constraints---in particular, constraints on the parent-$\Bsig$ (or equivalently -$\Btag$) flying {\em directions}, inferred from {\em vertexing} information.

A third novelty is the actual application, to namely $B$-meson decays, specifically to the $B^\pm \to K^\pm \tau \mu$ search at Belle and Belle II. This decay is of great interest at present, as it represents an expected further signature of many SM extensions explaining the coherent set of discrepancies in semi-leptonic $B$ decays known as ``$B$ anomalies''~\footnote{%
$B$ anomalies suggest new physics dominantly coupled to the third generation of down-type fermions \cite{Glashow:2014iga}. This, by way of flavour mixing after electroweak-symmetry breaking implies dominant (flavoured) effects in $b \to s$ transitions, and in final states with $\tau$s, including lepton-flavour violating ones \cite{Glashow:2014iga}. (For extensive formulae see \cite{Becirevic:2016oho}.) These observations were made properly $SU(2)_L$-symmetry compliant in Ref.~\cite{Bhattacharya:2014wla}, thus paving the way for joint explanations of $b \to s$ and $b \to c$ data (see also discussion in Ref.~\cite{Greljo:2015mma}). The limitations imposed by data on such simple picture, and paths to overcome such limitations, were discussed in Ref. \cite{Buttazzo:2017ixm}. One clear-cut direction is to consider a minimally-broken $U(2)^5$ global symmetry \cite{Barbieri:2011ci,Barbieri:2012uh}, that naturally addresses the resemblance between the seemingly hierarchical NP couplings to the three generations and the hierarchy in SM fermion masses.}.
The search strategy in place at $B$ factories (see e.g. Refs.~\cite{Belle_Note_1576,BaBar:2007jtd}) for the $B^\pm \to K^\pm \tau \mu$ decay is based on $\Bsig^+ \to \Ksig^+ \tau \ell_{\sig}$ events, with 1-prong $\tau$ decays to $\ell \nu \nu$, $\pi \nu$ and $\rho \nu$ (making up over 70\% of all $\tau$ decays), and associated with a {\em fully reconstructed, hadronic} tag decay, e.g. $\Btag^- \to D^0 (\to K^- \pi^+) \pi^-$~\footnote{Here we specified only correlated charge assignments. The search includes the case of flipped signs.}. This $\Btag$ decay is referred to as hadronic $B$-tagging, and allows to completely reconstruct the $\Bsig$ decay as well. There exists an entire set of additional, independent {\em semi-leptonic} (SL) tag decays such as $\Btag^- \to D^0 (K \pi) \ell^- \bar \nu$, which at present cannot be exploited within the above strategy. In fact, these decays involve escaping neutrinos, which hinder full event reconstruction. The strategy we describe makes events with SL tags competitive to events with hadronic tags, as we will show in terms of increase in statistics and of gain in branching-ratio sensitivity. Our application is meant to benchmark the strategy, and our results suggest it is exploitable in numerous other applications, some of which we mention at the end of the letter.

Within our chosen application, the overall resolution of a given set of decay channels---whether hadronic or SL---may be quantified by $\Mreco$. This has the crucial advantage of reducing the search to a `bump hunt' in the total invariant mass of the signal-side $\tau$ decay products.
$\Mreco$ can be constructed as
\begin{widetext}
\bea
\label{eq:Mrecoil}
M^2_{\rm recoil} \equiv (p^*_{e^+ e^-} - p^*_{\Btag} - p^*_{\Ksig \lsig})^2 = m^2_{\Btag} + m^2_{\Ksig \lsig} - 2 (E^*_{\Btag} E^*_{\Ksig \lsig} + |\p{\Btag}^*| |\p{\Ksig \lsig}^*| \cos \theta)~.
\eea
\end{widetext}
Here asterisks denote the center-of-mass frame, where $\Mreco$ takes a simple form, although it is Lorentz-invariant by definition; besides, $E^*_{\Btag} = \sqrt s /2$, and $\theta$ is the angle between $\p{\Btag}^*$ and $\p{\Ksig \lsig}^*$. Eq. (\ref{eq:Mrecoil}) allows to immediately identify the main current limitation of SL tags: since the tag side is not fully reconstructed, $\theta$ is unknown. For SL-tag analyses, the cosine of this angle is currently taken as uniformly distributed, and because of that the SL-tag current resolution is about 5 times worse than the hadronic tag's.

The above discussion allows to restate the task at hand as that of improving the $\cos\theta$ estimate in SL tags. To introduce our approach we first provide some more kinematic notation. We consider $B$ mesons pair-produced in electron-positron collisions at the $\Upsilon(4S)$ resonance
\begin{align}
\label{eq:topology}
e^+ e^- \to B_1 B_2 \to V_1(p_1) \chi_1(k_1) + V_2(p_2) \chi_2(k_2)~,
\end{align}
where $V_i$ are visible and $\chi_i$ invisible (sets of) particles. The center-of-mass energy $\sqrt{s}$ is fixed, and the transverse as well as the longitudinal momentum of the total $e^+ e^-$ system is known. The decay in eq.~(\ref{eq:topology}) lends itself to the construction of $\M2$~\cite{Barr:2011xt} (see also \cite{Ross:2007rm,Cho:2014naa}), i.e. the fully Lorentz-invariant extension of $\MT2$ \cite{Lester:1999tx,Barr:2003rg}. Given the decay topology, one may define $\M2$ in several different ways, according to the kinematic constraints---e.g. on-shell mass relations---that are imposed in the minimisation and those that are not~\cite{Cho:2014naa}.
Following customary notation, our general definition is
\begin{align}
\label{eq:M2c}
  M_{2 \mathcal C} =
  & \min_{\vb*{k}_{1}, \, \vb*{k}_{2} \in \mathbb{R}^3}
    \Big[ \max \Big\{ M(p_{1}, \, k_{1}), \, M(p_{2}, \, k_{2})
    \Big\} \Big]  \nonumber\\
  & \text{subject to}\,\,
    \begin{cases}
      \vb*{k}_{1} + \vb*{k}_{2} =
      \vb*{P}^\text{miss} , \\
      \mbox{more constraints} \to \mbox{fix } \mathcal C~,
    \end{cases}
\end{align}%
where $M$ denotes the invariant mass constructed from the sum of the momenta specified as arguments---note that these momenta correspond to those in eq.~(\ref{eq:topology}); the first constraint effectively reduces to three the d.o.f. over which the minimization is performed. The additional constraints in the bracket, to be discussed next, fix the subscript $\mathcal C$ in the variable's denomination.

A first on-shell constraint is provided by the total collision energy
\begin{align}
\label{eq:s}
(p_1 + k_1 + p_2 + k_2)^2 = s~,
\end{align}
that is known at Belle and Belle II; a further one is represented by the known masses of the decaying parents
\begin{align}
\label{eq:mB}
(p_1 + k_1)^2 = (p_2 + k_2)^2 = m_B^2~.
\end{align}
Note that the constraints (\ref{eq:s}) and (\ref{eq:mB}), taken together, reduce to {\em zero} the number of d.o.f. in the $\M2$ minimization---i.e. make the $M_2$ solution equivalent to finding the unique root of a complete set of kinematic equations for the event. A first $\M2$ definition will be eq.~(\ref{eq:M2c}) plus the constraints (\ref{eq:s}) and (\ref{eq:mB}), namely $M_{2sB}$. In the literature, the constraint in eq.~(\ref{eq:mB}) has been discussed {\em without} the last equality \cite{Konar:2015hea}, i.e. without assuming a known mass for the parent particles---in this case $\M2$ is known to have the same minimum as $M_{2s}$ \cite{Konar:2015hea}. To our knowledge (and surprise), a discussion of $\M2$ with the full constraint in eq.~(\ref{eq:mB}) is missing in the literature. This may be due to various reasons. $\M2$, as a generalization of $\MT2$, was born as a variable for direct searches of new---hence with unknown mass---pair-produced resonances decaying semi-invisibly; in fact, one of the defining features of $\M2$ is that its endpoint as a distribution allows to measure the decaying-parent's mass. Besides, the constraint in eq.~(\ref{eq:mB}), when imposed on top of eq.~(\ref{eq:s}), closes the event kinematics. The thus-obtained $\M2$ definition, to be referred to as $M_{2sB}$ in the numerical analysis, is thereby no more a distribution---its minimum is an exact solver of the kinematic equations, event by event.

The constraints hitherto discussed are `standard' on-shell ones. We next consider a further, qualitatively different, class of constraints---on the flight direction of the parent particles. Such constraints may be valuable because of the accurate vertexing information available at Belle and Belle II (and elsewhere). Note that imposing e.g. the $\Bsig$ flight direction renders the $\Btag$ counterpart redundant \footnote{In the center-of-mass (CM) frame, $\p{\Btag} = -\p{\Bsig}$ and that, at Belle (II), the boost from the lab to the CM frame is known from $\p{\Upsilon(4S)}$. As a consequence, imposing the constraint (\ref{eq:dphi}) also on the tag side is redundant.}, so we will discuss the former only.

The $\Bsig$ flight direction is determined, event by event, as $\hat{\boldsymbol v}_\sig = (\boldsymbol r_\sig - \boldsymbol r_0) / |\boldsymbol r_\sig - \boldsymbol r_0|$, where $\boldsymbol r_0$ and $\boldsymbol r_\sig$ are the locations of the primary and respectively the $\Bsig$-decay vertices. In principle, since each component of $\boldsymbol r_{0}$ as well as $\boldsymbol r_{\sig}$ comes with an error, the $\hat{\boldsymbol v}_\sig$ constraint could be implemented as
\be
\label{eq:dphi}
\arccos (\hat{\boldsymbol p}_{B_\sig} \cdot \hat{\boldsymbol v}_\sig) \le \delta_{\sig}~,
\ee
where $\hat{\boldsymbol p}_{B_\sig}$ denotes the unit vector corresponding to $\p{B_\sig} = \p{\sig} + \k{\sig}$, with $\k{\sig}$ estimated through MAOS~\cite{Cho:2008tj,Park:2011uz}. This constraint would dictate that the $\Bsig$ direction form a cone of maximal aperture $2\delta_{\sig}$ with $\hat{\boldsymbol v}_\sig$. As the error on $\boldsymbol r_{0,\sig}$ tends to zero, so does $\delta_{\sig}$ \footnote{The inequality constraint in eq.~(\ref{eq:dphi}) may be naturally implemented using the sequential quadratic programming (SQP) method \cite{Wilson1963,Palomares1976,Han1976,Han1977,Powell1978}, which is the main algorithm in the {\tt YAM2} software library~\cite{Park:2020bsu} utilized throughout this work.}.
Note that eq. (\ref{eq:dphi}) amounts to an {\em inequality} constraint, whose application does not reduce the number of d.o.f. in the $\M2$ minimisation. Inequality constraints may be very powerful for the purpose of overloading $\M2$, namely of bounding its minimisation with a number of constraints that, were they equalities, would equal or exceed the number of kinematic d.o.f. available. We will return to this point later. In our circumstances, we can follow a simpler procedure. Event by event, we replace the true $\hat{\boldsymbol v}_\sig$ with a vector estimated by smearing with motivated distributions both $\boldsymbol r_{0}$ and $\boldsymbol r_{\sig}$ around their true values. We then impose the thus-estimated $\hat{\boldsymbol v}_\sig$ as an {\em equality} constraint (henceforth `$V$') on the $\Bsig$ flight direction.

As regards the $\boldsymbol r_{0, \sig}$ smearing, we note that $(\boldsymbol r_{0})_{x,y}$, namely the primary-vertex' components in the plane orthogonal to the $e^+$-beam axis $z$, are determined very accurately. The beam has a non-negligible size mostly in the $z$ direction. At Belle, the beam $z$-profile is $\sigma_{z}^{r_0} \sim 4\,$mm, which makes the $\hat{\boldsymbol v}_\sig$ constraint ineffectual~\footnote{Note that the $B$ decay length is about one o.o.m. smaller than $\sigma_{z}^{r_0}$.}. Instead, the current $\sigma_{z}^{r_0}$ figure at Belle II is 350$\,\mu$m, which is expected to further improve to 150$\,\mu$m at the design luminosity~\cite{Belle-II:2010dht}. The secondary vertex $\boldsymbol r_{\sig}$ is determined with a spread in each coordinate of about 45$\,\mu$m at Belle and 25$\,\mu$m at Belle II. In short, we estimate $\boldsymbol r_{0, \sig}$ as normally distributed random numbers, with standard deviations given by the spreads discussed.

`Directional' constraints such as eq.~(\ref{eq:dphi}) have, to our knowledge, never been considered in connection with $M_2$ and siblings. The $\M2$ definition of eq.~(\ref{eq:M2c}), plus the (equality) $V$ constraint just discussed will be referred to as $M_{2V}$ \footnote{We verified that $\M2$ plus the inequality constraint of eq.~(\ref{eq:dphi}) yields, in the small-$\delta_\sig$ limit, the same solution as the respective equality constraint. Needless to say, imposing an equality constraint is however faster and less subject to numerical instabilities.}.
$M_{2V}$ reduces the number of d.o.f. available in the $\M2$ minimization by two units---the constraint fixes one direction in 3-dimensional space.
Note that, if we further use the knowledge of $s$, we can fix not only the orientation, but also the magnitude of $\p{\Bsig}$, and thereby close the kinematics. In this case, that we will denote as $M_{2sV}$, the minimization must necessarily land at $\k{\sig} = \k{\sig}^\true$. On the other hand, if one does {\em not} use the $s$ information, as in $M_{2V}$, $s$ is an outcome of the $\M2$ algorithm.

Before turning to the discussion of $\Mreco$ and of the ensuing $\mc B(B \to K \tau \mu)$ limit from $\M2$, two points are in order. The first concerns our numerical setup and assumptions. Our results use phase-space events populated through the {\tt EvtGen} Monte Carlo \cite{Lange:2001uf}. For consistency with the on-going Belle analysis \cite{Belle_Note_1576}, and as already mentioned, we restrict to the 1-prong $\tau$ decays to $\ell \nu \nu$, $\pi \nu$ and $\rho \nu$. We note that our thus-generated event sample includes realistic detector smearing, which in our case affects $\p{i}$ (see eq. (\ref{eq:topology})), ${\boldsymbol P}^{\miss}$, and the locations of the interaction point and of the decay vertices. We also assume that the combinatorial ambiguity due to the assignment of the visible particles to one of the two decay chains has completely been resolved. The purpose is to single out the signal quality degradation due {\em solely} to the presence of undetected particles on the tag side.

\begin{figure}[b]
  \begin{center}
  \vspace{-0.3cm}
    \includegraphics[width=0.238\textwidth]{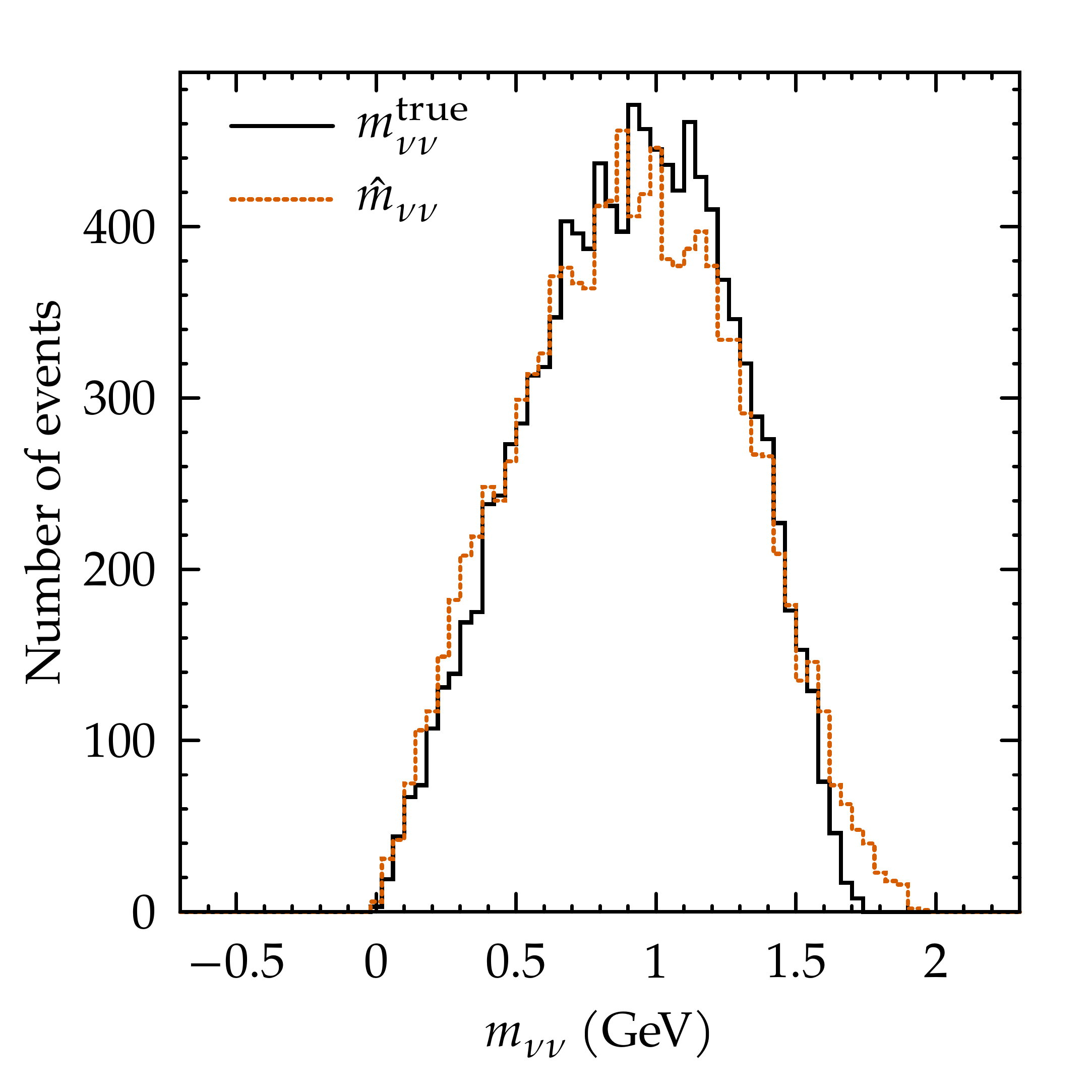}
    \includegraphics[width=0.238\textwidth]{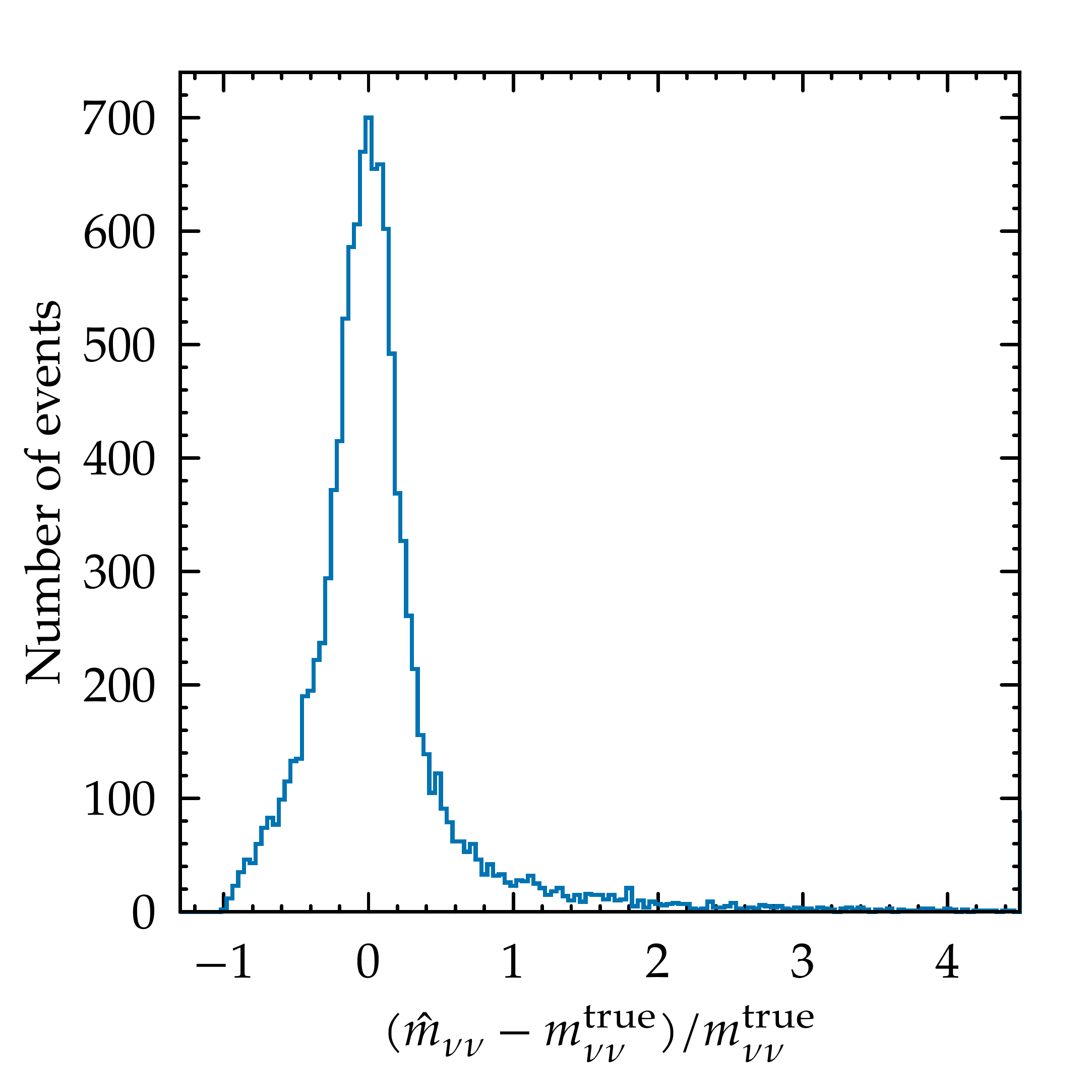}
  \end{center}
  \vspace{-0.4cm}
  \caption{Distribution for $\hat m_{\nu\nu}$ as defined in the text, and comparisons with the truth-level distribution.
\label{fig:Mnunu}
}
\end{figure}

The second point is the following. The constraints (\ref{eq:s}) and (\ref{eq:mB}) require knowledge of $k_{1,2}^2$, the invariant masses of the invisible systems on the signal and tag sides. While $k_2^2 = 0$, $k_1^2$ is unknown for a leptonically-decaying signal-side $\tau$---it is the invariant mass squared of the two final-state neutrinos. A simple ansatz often adopted in the literature is $k_1^2 = 0$. This is however not realistic, as the truth-level $m_{\nu \bar \nu}^2$ distribution peaks around $1~\GeV^2$, as shown in Fig. \ref{fig:Mnunu} (left).
We then consider an improved ansatz for $k_1^2$, expecting that such improvement will play a role mostly in the $\M2$ constraints. In fact, $k_1^2$ enters also the very definition of $\M2$, but existing literature suggests that this dependence is of lesser consequence \cite{Cho:2008tj, Park:2011uz, Guadagnoli:2021fcj}. We construct our $k_1^2$ ansatz as follows. We start from the $B$ in its rest frame, and boost it with the total beams momentum. We then obtain $k_1$ by subtracting the sum of the visible final-state momenta. This approximation neglects the back-to-back momentum of the $B - \bar B$ pair in the center-of-mass frame, which is however small with respect to the boost induced by the beams asymmetry. As a result of our ansatz, the $k_1^2$ distribution is neatly close to the truth-level one for about 86\% of the events, and yields unphysical negative values for the remaining 14\%. For these events we switch the overall $k_1^2$ sign. The resulting ansatz for the $\nu \bar \nu$ invariant mass, to be denoted as $\mvva$, gives rise to the distribution shown in Fig. \ref{fig:Mnunu}.
\begin{figure*}[t]
  \begin{center}
    \includegraphics[width=0.9\textwidth]{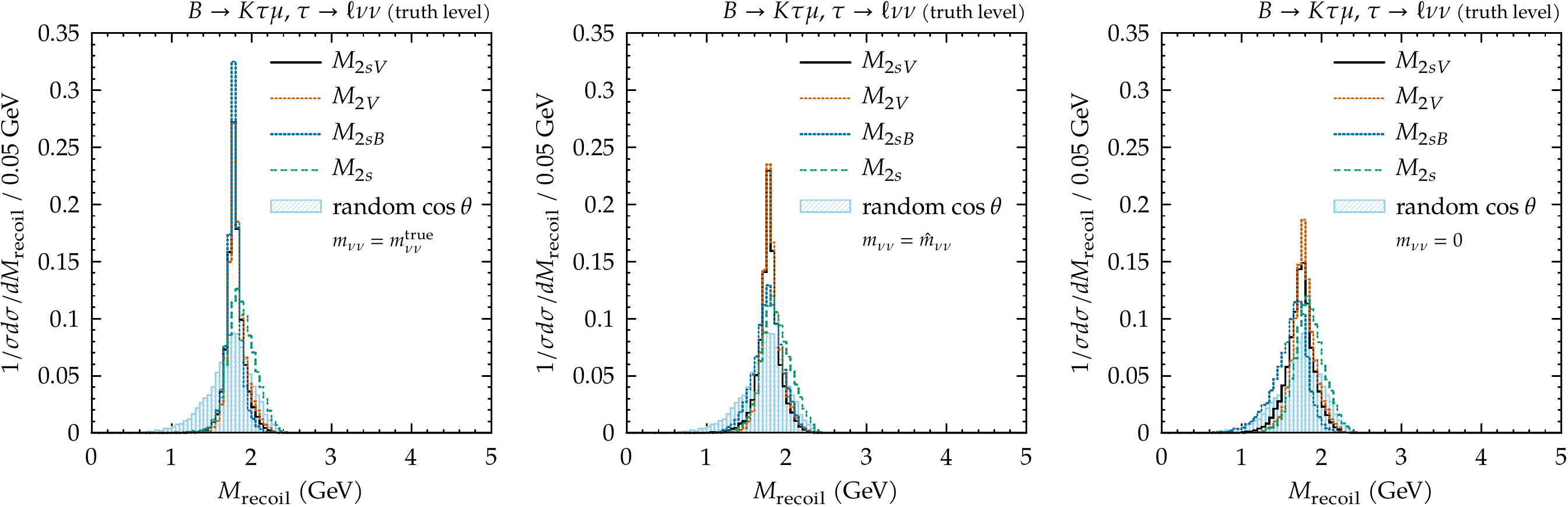}
  \end{center}
  \vspace{-0.50cm}
  \caption{$\Mreco$ distributions for $M_{2s(B)}$ and $M_{2(s)V}$ (see lines in each plot's legend), as a function of $\mvv = \{\mvv^{\rm true}$, $\mvva$, $0\}$ (panels left to right). The random-$\cos\theta$ distribution is also shown as reference.
\label{fig:Mnunu_true_vs_approx_vs_0}
}  \vspace{-0.20cm}
\end{figure*}

We implemented $\mvva$ in all of $M_{2s(B)}$ and $M_{2(s)V}$ and tested the improvement directly in the $\Mreco$ distribution. In Fig. \ref{fig:Mnunu_true_vs_approx_vs_0} we show these four $\M2$ definitions, plus the random-$\cos\theta$ distribution as reference. The three panels correspond to the three choices $\mvv = \{\mvv^{\rm true}$, $\mvva$, $0\}$, respectively. This figure prompts the following comments. First, in terms of the distributions' height/width, the improvement due to $\mvva$ with respect to $\mvv = 0$ is significant for the $M_{2(s)V}$, whereas it is only slight for $M_{2sB}$. However, for $M_{2sB}$, the peak position with the $\mvv = 0$ ansatz is lower than the correct value, and this bias disappears with the $\mvva$ ansatz. All things considered, the accuracy of the $\mvv$ ansatz appears to impact especially $M_{2sB}$. This sensitivity seems to be due to the $B$ constraint, because the quality of the peak in $\Mrecoil$ calculated with $M_{2sB}$ improves dramatically from the $\mvva$ to the $\mvv^{\rm true}$ cases, whereas the corresponding improvement in $M_{2s}$ or $M_{2(s)V}$ is marginal or absent.

\begin{figure*}[t]
  \begin{center}
    \includegraphics[width=0.90\textwidth]{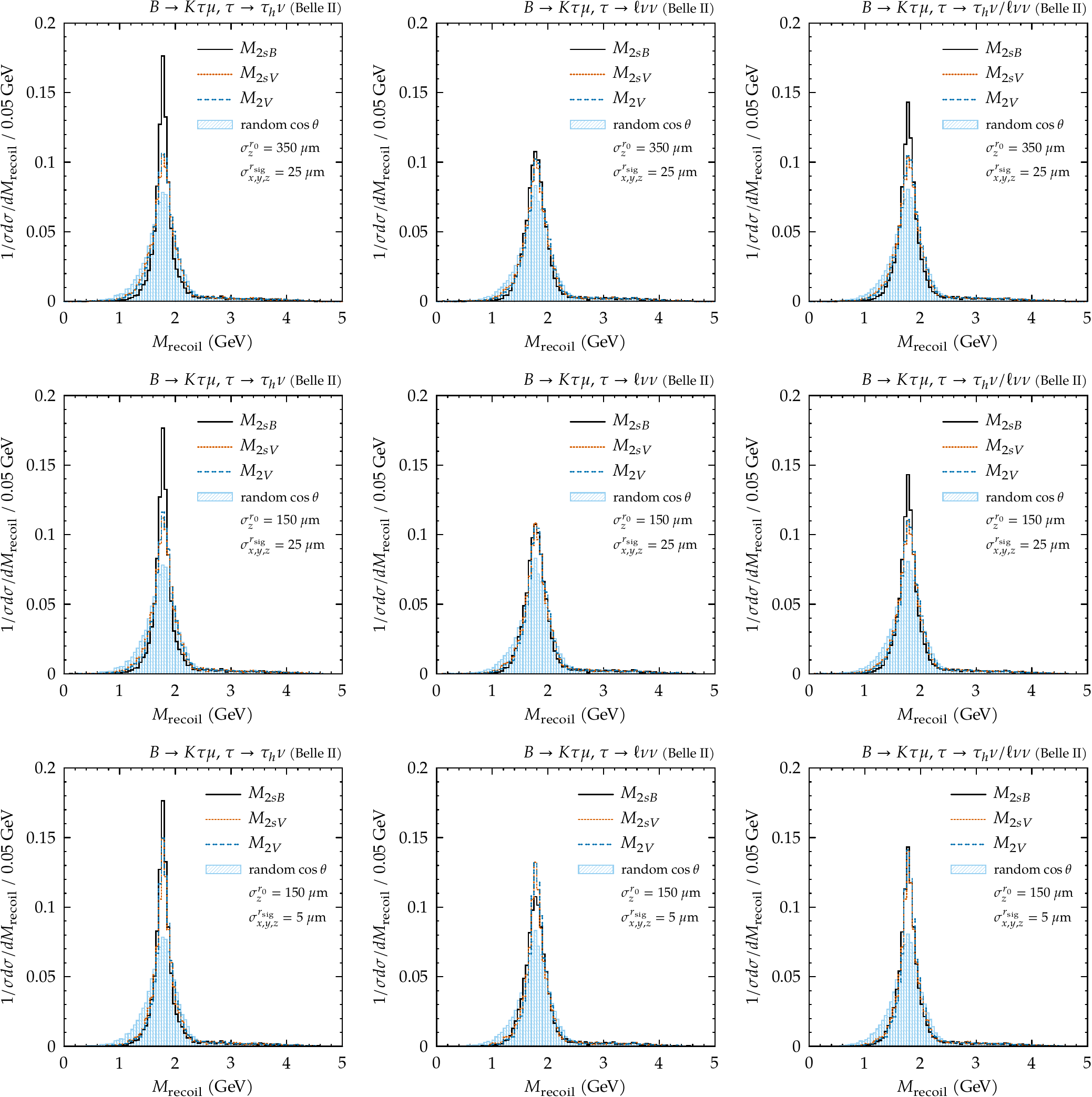}
  \end{center}
  \vspace{-0.3cm}
  \caption{$\Mreco$ distributions for different experimental setups, detailed in the text, and for different $\tau_\sig$ decay modes (leftmost to rightmost panels: hadronic, leptonic $\tau$ decays, or both).
\label{fig:M2sB_vs_M2starV_in_Mreco}
} \vspace{-0.4cm}
\end{figure*}
We next compare the performance of $M_{2sB}$, $M_{2V}$ and $M_{2sV}$ in the $\Mreco$ distribution. This comparison is presented in the histogram array of Fig. \ref{fig:M2sB_vs_M2starV_in_Mreco}. The performance of the random-$\cos\theta$ case is again shown as baseline. To ease readability, we note that the array's columns refer to the different 1-prong decay modes considered for the signal tau: from left to right, hadronic, leptonic $\tau$ decays, or both. The choice of the channels affects all of $M_{2sB}$, $M_{2V}$ and $M_{2sV}$. The array's rows, in turn, represent different scenarios for $\sigma_z^{r_0}$ and $\sigma_{x,y,z}^{r_{\sig}}$: the first row, with $\sigma_z^{r_0} = 350\,\mu$m and $\sigma_{x,y,z}^{r_{\sig}}= 25\,\mu$m, represents the current setup at Belle II; the same $\sigma_{x,y,z}^{r_{\sig}}$ value is used in the second row, along with $\sigma_z^{r_0} = 150\,\mu$m, that represents the setup at the Belle II design luminosity, as discussed above. Finally, in the lowest row
$\sigma_z^{r_0}$ stays at this value, whereas $\sigma_{x,y,z}^{r_{\sig}}$ are decreased to the hypothetical value of $5\,\mu$m, 
for reasons to be discussed shortly. By definition, $\sigma^{r_0}$ or $\sigma^{r_\sig}$ affect $M_{2(s)V}$ only, not $M_{2sB}$. The $M_{2sB}$ distributions are shown in every figure row only for comparison with the respective $M_{2(s)V}$ distributions.

This comparison shows that $M_{2sB}$ performs better than $M_{2(s)V}$ in the hadronic $\tau_\sig$-decay case, whereas $M_{2sB}$ and $M_{2(s)V}$ are comparable in the leptonic-decay instance. This implies a somewhat better $M_{2sB}$ performance when the channels are combined. These conclusions hold in the ``current Belle-II'' scenario (first row) and to a lesser degree in the ``Belle-II design-luminosity'' scenario (second row). We see from the lower two rows that, if the $\sigma^{r_\sig}$ accuracy were to halve with respect to the 25$\,\mu$m figure, the $M_{2(s)V}$ performance would be very close to $M_{2sB}$ in the hadronic $\tau_\sig$-decay case, and even superior to it in the leptonic-decay case, implying a comparable performance between $M_{2sB}$ and $M_{2(s)V}$ in the combined-channel case.

These findings suggest that $M_{2s(B)}$ and $M_{2(s)V}$ have distinct advantages and disadvantages: the former has a strong sensitivity to the $\mvv$ constraint; the latter has little sensitivity in that respect, and allows to profitably use vertexing information---to the extent that the latter is accurate enough. In fact, the choice of $M_{2(s)V}$ over $M_{2sB}$ hinges on the considered detector's vertexing capabilities, and in case of comparable performances the best strategy would be a combined analysis, where $\M2$ is overloaded with all of the $s$, $B$ and $V$ constraints.

The $\Mreco$ distributions can finally be translated into an upper limit on $\mc B(B \to K \tau \mu)$. This is calculated at 90\% confidence level with an established frequentist method (see e.g. Ref. \cite{BELLE:2019xld}). For each given $\Mreco$ distribution, we apply our selection to a Monte Carlo sample consisting of all possible backgrounds, of overall size equal to the Belle dataset. This allows to constrain the background shape in the signal region beyond a simple sideband extrapolation.
Given the superiority of $M_{2sB}$ over $M_{2(s)V}$ within the Belle-II setup in the foreseeable future, for this study we deploy $M_{2sB}$ alone, with the $\mvva$ ansatz for $\tau_\sig \to \{\rho,\pi\} + \nu$. The analysis uses somewhat simplifying assumptions: 1-prong $\tau_\sig$ decays are reconstructed as $\pi / \mu / e$, i.e. the $\rho$ is currently not being reconstructed; besides cross-feed across the different categories is neglected. These effects, however, are not expected to sizeably change the overall picture. With these simplifications, we get a 90\% CL upper bound on $\mc B (B^\pm \to K^\pm \tau^\pm \mu^\mp) = 1.2 \times 10^{-5}$ with $710~{\rm fb}^{-1}$. Within our approximations, this limit equals the one that we obtain with the hadronic tag. The corresponding limits obtained with $\Mreco$(random $\cos \theta$) and with $\Mreco^\true$ are respectively $2.0$ and $0.6$ both in units of $10^{-5}$. Hence a no-frills application of $M_{2sB}$ leads per se to an improvement already halfway between the current strategy and the fully-reconstructed SL case.

Our approach opens several lines of development. First, the constraints discussed can be applied to $\M2$, eq.~(\ref{eq:M2c}), in a number of combinations, according not only to the constraints that are actually included, but also to whether they are imposed exactly (i.e. as equalities) or as inequality relations. In particular, one may exceed the total number of kinematic d.o.f. available---thus overloading $\M2$---by implementing constraints as inequalities. In this respect, $V$ constraints are especially promising, also in that they can be deployed at hadronic facilities, where $s$ is not available, but accurate vertexing typically is. Second, the $B \to K \tau \mu$ analysis we discussed may be carried over to many additional channels of topical interest. One example is $B \to \tau \mu$, that features at least two clear advantages: the $\tau$ is monochromatic in the parent-$B$ rest frame, and the signal has lower combinatorial backgrounds. Another example is $B \to K \nu \bar \nu$, a key constraint for model-building of the current $B$ anomalies \cite{Buras:2014fpa} and even for searches of light supersymmetric states (e.g. \cite{Dib:2022ppx}). We expect our approach to make SL tags as practicable as hadronic tags in these and other modes. Our work is in the direction of what may be denoted as the ``efficiency frontier''. The next step towards improving tag-based analyses even further is then to enlarge the set of hadronic tags used --- in what is instead the ``resolution frontier''. This step seems however very dependent on the specific decay considered, i.e it does not seem to admit a universal approach such as the one laid out here.

This work is supported by ANR under contract n. 202650 and by IBS under the project code IBS-R018-D1.

\bibliography{bibliography}

\end{document}